\RequirePackage{lineno}

\documentclass[preprint, 12pt]{aastex}
\usepackage{amsmath}
\usepackage{natbib}
\begin{document}
\linenumbers

\title{Multiwavelength Observations of The TeV Binary LS I +61$^{\circ}$ 303 with VERITAS, Fermi-LAT and Swift-XRT During a TeV Outburst}

\author{
E.~Aliu\altaffilmark{1}
S.~Archambault\altaffilmark{2},
B.~Behera\altaffilmark{3},
K.~Berger\altaffilmark{4},
M.~Beilicke\altaffilmark{5},
W.~Benbow\altaffilmark{6},
R.~Bird\altaffilmark{7},
A.~Bouvier\altaffilmark{8},
V.~Bugaev\altaffilmark{5},
M.~Cerruti\altaffilmark{6},
X.~Chen\altaffilmark{3,9},
L.~Ciupik\altaffilmark{10},
M.~P.~Connolly\altaffilmark{11},
W.~Cui\altaffilmark{12},
J.~Dumm\altaffilmark{13},
A.~Falcone\altaffilmark{14},
S.~Federici\altaffilmark{3,9},
Q.~Feng\altaffilmark{12},
J.~P.~Finley\altaffilmark{12},
P.~Fortin\altaffilmark{6},
L.~Fortson\altaffilmark{13},
A.~Furniss\altaffilmark{8},
N.~Galante\altaffilmark{6},
G.~H.~Gillanders\altaffilmark{11},
S.~Griffin\altaffilmark{2},
S.~T.~Griffiths\altaffilmark{15},
J.~Grube\altaffilmark{10},
G.~Gyuk\altaffilmark{10},
D.~Hanna\altaffilmark{2},
J.~Holder\altaffilmark{4},
G.~Hughes\altaffilmark{3},
T.~B.~Humensky\altaffilmark{1},
P.~Kaaret\altaffilmark{15},
M.~Kertzman\altaffilmark{16},
Y.~Khassen\altaffilmark{7},
D.~Kieda\altaffilmark{17},
F.~Krennrich\altaffilmark{18},
M.~J.~Lang\altaffilmark{11},
G.~Maier\altaffilmark{3},
P.~Majumdar\altaffilmark{19,20},
S.~McArthur\altaffilmark{21},
A.~McCann\altaffilmark{22},
P.~Moriarty\altaffilmark{23},
R.~Mukherjee\altaffilmark{1},
A.~O'Faol\'{a}in de Bhr\'{o}ithe\altaffilmark{7},
R.~A.~Ong\altaffilmark{19},
A.~N.~Otte\altaffilmark{24},
N.~Park\altaffilmark{21},
J.~S.~Perkins\altaffilmark{25},
M.~Pohl\altaffilmark{3,9},
A.~Popkow\altaffilmark{19},
H.~Prokoph\altaffilmark{3},
J.~Quinn\altaffilmark{7},
K.~Ragan\altaffilmark{2},
J.~Rajotte\altaffilmark{2},
G.~Ratliff\altaffilmark{10},
P.~T.~Reynolds\altaffilmark{26},
G.~T.~Richards\altaffilmark{24},
E.~Roache\altaffilmark{6},
G.~H.~Sembroski\altaffilmark{12},
F.~Sheidaei\altaffilmark{17,*},
C.~Skole\altaffilmark{3},
A.~W.~Smith\altaffilmark{17,*},
D.~Staszak\altaffilmark{2},
I.~Telezhinsky\altaffilmark{3,9},
J.~Tyler\altaffilmark{2},
A.~Varlotta\altaffilmark{12},
S.~Vincent\altaffilmark{3},
S.~P.~Wakely\altaffilmark{21},
T.~C.~Weekes\altaffilmark{6},
A.~Weinstein\altaffilmark{18},
R.~Welsing\altaffilmark{3},
A.~Zajczyk\altaffilmark{5}
B.~Zitzer\altaffilmark{27}
}
\altaffiltext{*}{Corresponding Authors: aw.smith@utah.edu, sheidaei@physics.utah.edu}
\altaffiltext{1}{Physics Department, Columbia University, New York, NY 10027, USA}
\altaffiltext{2}{Physics Department, McGill University, Montreal, QC H3A 2T8, Canada}
\altaffiltext{3}{DESY, Platanenallee 6, 15738 Zeuthen, Germany}
\altaffiltext{4}{Department of Physics and Astronomy and the Bartol Research Institute, University of Delaware, Newark, DE 19716, USA}
\altaffiltext{5}{Department of Physics, Washington University, St. Louis, MO 63130, USA}
\altaffiltext{6}{Fred Lawrence Whipple Observatory, Harvard-Smithsonian Center for Astrophysics, Amado, AZ 85645, USA}
\altaffiltext{7}{School of Physics, University College Dublin, Belfield, Dublin 4, Ireland}
\altaffiltext{8}{Santa Cruz Institute for Particle Physics and Department of Physics, University of California, Santa Cruz, CA 95064, USA}
\altaffiltext{9}{Institute of Physics and Astronomy, University of Potsdam, 14476 Potsdam-Golm, Germany}
\altaffiltext{10}{Astronomy Department, Adler Planetarium and Astronomy Museum, Chicago, IL 60605, USA}
\altaffiltext{11}{School of Physics, National University of Ireland Galway, University Road, Galway, Ireland}
\altaffiltext{12}{Department of Physics, Purdue University, West Lafayette, IN 47907, USA }
\altaffiltext{13}{School of Physics and Astronomy, University of Minnesota, Minneapolis, MN 55455, USA}
\altaffiltext{14}{Department of Astronomy and Astrophysics, 525 Davey Lab, Pennsylvania State University, University Park, PA 16802, USA}
\altaffiltext{15}{Department of Physics and Astronomy, University of Iowa, Van Allen Hall, Iowa City, IA 52242, USA}
\altaffiltext{16}{Department of Physics and Astronomy, DePauw University, Greencastle, IN 46135-0037, USA}
\altaffiltext{17}{Department of Physics and Astronomy, University of Utah, Salt Lake City, UT 84112, USA}
\altaffiltext{18}{Department of Physics and Astronomy, Iowa State University, Ames, IA 50011, USA}
\altaffiltext{19}{Department of Physics and Astronomy, University of California, Los Angeles, CA 90095, USA}
\altaffiltext{20}{Saha Institute of Nuclear Physics, Kolkata 700064, India}
\altaffiltext{21}{Enrico Fermi Institute, University of Chicago, Chicago, IL 60637, USA}
\altaffiltext{22}{Kavli Institute for Cosmological Physics, University of Chicago, Chicago, IL 60637, USA}
\altaffiltext{23}{Department of Life and Physical Sciences, Galway-Mayo Institute of Technology, Dublin Road, Galway, Ireland}
\altaffiltext{24}{School of Physics and Center for Relativistic Astrophysics, Georgia Institute of Technology, 837 State Street NW, Atlanta, GA 30332-0430}
\altaffiltext{25}{N.A.S.A./Goddard Space-Flight Center, Code 661, Greenbelt, MD 20771, USA}
\altaffiltext{26}{Department of Applied Physics and Instrumentation, Cork Institute of Technology, Bishopstown, Cork, Ireland}
\altaffiltext{27}{Argonne National Laboratory, 9700 S. Cass Avenue, Argonne, IL 60439, USA}

\begin{abstract}

We present the results of a multiwavelength observational campaign on the TeV binary system LS~I~+61$^{\circ}$~303 with the VERITAS telescope array ($>$200 GeV), Fermi-LAT (0.3-300 GeV), and Swift-XRT (2-10 keV). The data were taken from December 2011 through January 2012 and show a strong detection in all three wavebands. During this period VERITAS obtained 24.9 hours of quality selected livetime data in which LS I +61$^{\circ}$ 303 was detected at a statistical significance of 11.9$\sigma$. These TeV observations show evidence for nightly variability in the TeV regime at a post-trial significance of 3.6$\sigma$. The combination of the simultaneously obtained TeV and X-ray fluxes do not demonstrate any evidence for a correlation between emission in the two bands. For the first time since the launch of the Fermi satellite in 2008, this TeV detection allows the construction of a detailed MeV-TeV spectral energy distribution from LS I +61$^{\circ}$ 303. This spectrum shows a distinct cutoff in emission near 4 GeV, with emission seen by the VERITAS observations following a simple power-law above 200 GeV. This feature in the spectrum of LS I +61$^{\circ}$ 303, obtained from overlapping observations with Fermi-LAT and VERITAS, may indicate that there are two distinct populations of accelerated particles producing the GeV and TeV emission.

\end{abstract}

\section{Introduction}

The high-mass X-ray binary LS~I~+61$^{\circ}$~303 is perhaps the most studied member of a surprisingly small class of X-ray binary systems which are also known sources of TeV emission. Despite many years of observations across the electromagnetic spectrum, the system remains, in some respects, poorly characterized. Known to be
the pairing of a massive B0 Ve star and a compact object of unknown nature \citep{Casares2005, HandC1981},
LS~I~+61$^{\circ}$~303 has been known historically for its energetic outbursts at radio, X-ray, GeV, and TeV wavelengths \citep{Abdo2009,VERITASLSIDetection,Albert2006,Gregory2002,GR2001,Harrison2000, Zhang2010}, all of these showing correlation with the 26.5 day orbital cycle of the compact object. Radial velocity measurements show the orbit to be elliptical ($e=0.537\pm0.034$),
with periastron passage occurring around phase $\phi=0.275$, apastron passage at $\phi=0.775$, superior conjunction at $\phi=0.081$ and inferior conjunction at $\phi=0.313$ \citep{Aragona2009}. Although it should be noted that all of the orbital parameters of LS I +61$^{\circ}$ 303 are subject to some uncertainty as the inclination of the system is not precisely known.

Observations in the non-thermal regime have managed to illustrate some key phenomena. Extensive observations by both RXTE and Swift-XRT have provided a wealth of X-ray data which show a regular emission period consistent with the orbital period \citep{Smith2008, Esposito2007}.  The modulation of this X-ray peak is seen on multiple timescales, from individual orbits up to several years; most importantly, a modulation on a $\sim$4.5 year timescale \citep{LiXray, Chernya2012} has been observed in the hard X-ray band, reminiscent of the well known 4.5 year modulation of the radio period \citep{Gregory2002}. However, a definitive link between the particle acceleration processes producing the radio emission and those producing the X-ray emission is still lacking. An additional feature of the system is the possible association of short ($<$0.1s), high luminosity X-ray bursts from the system \citep{Pasquale2008, Burrows2012} which have been interpreted as the result of the emission from a high magnetic field neutron star \citep{Papitto2012}. Further observations of such behavior from the system in the X-ray band, definitively linked to LS I +61$^{\circ}$ would solidify this association.

In the GeV band, LS I +61$^{\circ}$ 303 was one of the few non-pulsar galactic objects firmly identified in the initial Fermi-LAT Bright Source List with an average flux of $(0.82\pm0.03_{stat} \pm 0.07_{syst}) \times 10^{-6}$ $\gamma$ cm$^{-2}$ s$^{-1}$ above 100 MeV \citep{Abdo2009}. The spectrum showed an exponential cutoff at 6.3 $\pm$ 1.1$_{stat}$ $\pm$ 0.4$_{sys}$ GeV and a photon index of $\Gamma$ = 2.21 $\pm$ 0.04$_{stat}$ $\pm$ 0.06$_{sys}$. In the first 8 months of LAT data, the source demonstrated a clear modulation of GeV emission with a period of $\sim$26.5 days, compatible with the radio period. The highest GeV fluxes were measured around phase $\phi$ = 0.4, close to periastron. However, subsequent analysis of $\sim$4.5 years of Fermi-LAT data shows clear evidence for long term variability of the mean orbital flux along with the apparent disappearance of its previously observed orbital modulation \citep{Hadasch}. This long term variability has recently been elucidated in \citet{Fermi2013}, where the Fermi-LAT collaboration shows a detection of the $\sim$4.5 year modulation of the GeV flux around apastron, consistent with the modulation seen in both radio and X-rays. 

As a TeV source, the system has presented puzzling behavior. Initial detections in 2006-2007 by both the VERITAS and MAGIC collaborations \citep{Albert2006, VERITASLSIDetection} over many orbital cycles showed the source to be a variably bright TeV source, with emission peaking around apastron passage. Subsequent observations in 2008-2010 \citep{Acciari2010} showed no evidence for emission during these previously detected phases, instead only detecting the source at a lower TeV flux near the periastron passage of a single orbit. The connection between the observed emission in different energy bands is not clear; initial detections of a correlation between the TeV and X-ray fluxes \citep{Albert2008} were not seen in later observations. Additionally, previous observations have not shown the GeV and TeV emission from the system to be strongly correlated either \citep{Acciari2010}. 

 As is the case with many TeV sources, the models to explain observed emission consist of both leptonic (inverse Compton scattering) and hadronic (pion decay resulting from relativistic proton interactions) variations. LS I +61$^{\circ}$ 303 is certainly no different in this respect, however, the confusion between emission models is compounded by an ambiguity in what type of engine actually powers the particle acceleration. LS I +61$^{\circ}$ 303 was originally thought to be a microquasar system due to the observation of what appeared to be extended radio jets \citep{Massi2001}. In this scenario, emission from the system is powered by a variably fed accretion disk which, in turn, powers a relativistic jet. The variability observed across the spectrum would then be explained by the accretion disk's exposure to varying levels of the strong stellar wind common to Be star systems. This model (under the assumption of basic Bondi-Hoyle accretion) would then predict non-thermal emission in the various bands to be coupled (in the simplest scenario) with the maximum flux occurring near periastron passage where the density of the stellar material is greatest. While this appears to be true sometimes in the GeV regime, it is not true in the TeV regime where emission is typically at a maximum near apastron passage. 
 
 However, the existence of a radio jet (and the validity of using a microquasar scenario) was called into question by high resolution VLBA imaging which observed what appeared to be the cometary emission from the interaction between a pulsar wind and the wind of the stellar companion \citep{Dhawan2006}. In this scenario (where the emission is powered by a shock front between the two winds) the variability would also result from varying levels of stellar wind density. However, in this model the emission in the various bands is decoupled by both the magnetic field strength at the shock and ``stand off distance" (distance from the shock to the pulsar)  changing as function of orbit. The change in these two quantities would dictate both cooling mechanisms and acceleration parameters, thus changing the relative intensities of emission between bands.
 
 It should be pointed out that neither of these models can explain all of the observed emission variability in the system (for instance, the VERITAS detection of TeV emission far away from apastron passage).  Additionally, since neither pulsations nor an accretion-like X-ray spectrum have yet to be observed in the system, current observations have not yielded a definitive answer to whether the system harbors a pulsar or black hole and both theoretical frameworks used to describe this system are still lacking strong constraints. What is clear however, is that the simplest version of either model will not adequately explain the observations. For example, both photon-photon absorption and line-of-sight effects almost certainly have to be taken into account when accounting for the observed variability. For examples of more recently advanced observations and models, see \citet{Zabalza}, \citet{newTorres} (binary pulsar model) and \citet{ZimmermanMassi} (microquasar model).
 
 Determining the correct physical model for this source requires additional dedicated observations across the multiwavelength spectrum.  In this work we detail the multiwavelength campaign on LS I +61$^{\circ}$ 303 incorporating both contemporaneous and simultaneous observations in the X-ray (Swift-XRT), GeV (Fermi-LAT), and TeV (VERITAS) regimes. This campaign was taken during a relatively strong period of emission in the TeV regime, and stands as the first time that  simultaneous GeV/TeV have been available during a high TeV state. During this high state, VERITAS detected marginal evidence for nightly variability in the system as well as a lack of strong correlated emission between the TeV flux and X-ray/GeV fluxes. Additionally, the spectral energy distribution obtained during these observations reveals a puzzling lack of detected emission between 30 and 200 GeV which makes the characterization of the gamma-ray emission non-trivial.

\section{VERITAS Observations}

The VERITAS array \citep{VERITAS} of imaging atmospheric Cherenkov telescopes (IACTs), located in southern Arizona (1.3 km a.s.l., 31$^{\circ}$40'30''N, 110$^{\circ}$57'07'' W), began 4-telescope array observations in September 2007. The array is composed of four 12m diameter telescopes, each with a Davies-Cotton tessellated mirror structure of 345 12m focal length hexagonal mirror facets (total mirror area of 110 m$^{2}$). Each telescope focuses Cherenkov light from particle showers onto its 499-pixel  photomultiplier tube camera. Each pixel has a field of view of 0.15$^{\circ}$, resulting in a camera field of view of 3.5$^{\circ}$. VERITAS has the capability to detect and measure gamma rays in the 100 GeV to 30 TeV energy regime with an energy resolution of 15-20$\%$ and an angular resolution of $<$0.1$^{\circ}$ on an event by event basis. 

VERITAS observed LS I +61$^{\circ}$ 303 beginning in early December 2011 (MJD 55911) until late January 2012 (MJD 55497), acquiring a total of 24.5 hours of quality selected, live-time observations. These observations provided detailed (although uneven) sampling of the phase bins $\phi$=0.45-0.05 of the binary orbit. Figure 1 shows the source light curve binned by both MJD and orbital phase. During the orbital phase regions of 0.5-0.8, the source was highly active, presenting a flux of 5-15 $\times 10^{-12} \gamma$s cm$^{-2}$s$^{-1}$ above 350 GeV, or approximately 5-15$\%$ of the Crab Nebula flux in the same energy regime. For the entire 24.5 hour observation, VERITAS detected an excess of 791 events from LS I +61$^{\circ}$ 303, equivalent to a detection at the 11.9$\sigma$ significance level. The data are used to create a differential energy spectrum from 0.2-5 TeV which is reasonably fit by a power-law ($\chi{2}$/n.d.f = 1.1/5) described by $(1.37\pm0.14_{stat})$$\times$10$^{-12}$ $\times$($\frac{E}{\mathrm{1 TeV}})^{-2.59\pm0.15_{stat}}$ $\gamma$s TeV$^{-1}$ cm$^{-2}$ s$^{-1}$. A comparison of this spectrum with those obtained by previous measurements at different flux levels \citep{Aleksic, VERITASLSIDetection} shows no indication for variability in the spectral slope from the source across a wide range of flux levels (see Figure 2).

The observations of LS I +61$^{\circ}$ 303 taken in 2011 also display an indication that the source may be variable in the TeV regime on a timescale much shorter than previously observed. While LS I +61$^{\circ}$ 303 is known to be a variable TeV source on the timescale of a single orbital period, the 2011 VERITAS observations indicate that the source may be variable on a nightly timescale. To test this hypothesis, we proceed by collecting the nightly absolute fluxes (Figure 1) and finding the pairs of observations which are separated by 1 day. Nine such pairs of observations exists within the 2011 observations and their fluxes are shown in Table 1. To test for variability on a nightly timescale we choose to test against the null hypothesis that, given a pair of nightly separated fluxes  (F$_{1}$, F$_{2}$),  F$_{2}$ was significantly larger than F$_{1}$ (as well as the inverse hypothesis). Assuming that both the source fluxes and errors are normally distributed, we construct the 2-dimensional Gaussian function:

\begin{equation} 
G(x,y) = \frac{1}{2 \pi \sigma_{1} \sigma_{2} }  e^{-\frac{(x-F_{1})^{2}}{2\sigma_{1}^{2}} - \frac{(y-F_{2})^{2}}{2\sigma_{2}^{2}}}            
 \end{equation}

where $\sigma$ represents the errors on the measured fluxes, and $x$ and $y$ are both flux space variables. Within this parametrization, a constant flux from night to night is represented by the function y=x. The probabilities that F$_{2}$ was greater than F$_{1}$ (or vice versa) can then be obtained by examining the integral:

\begin{equation}
\int_{-\infty}^{+\infty}dx\int_{x}^{+\infty}G(x,y)dy
\end{equation}

The resulting probabilities for F$_{1} >,<$F$_{2}$ (Table 1) show marginal evidence that the source is variable on a nightly timescale. The observations taken on MJD 55918/55919 and MJD 55944/55945 show evidence for a flux decrease at the 2.7$\sigma$ and 3.6$\sigma$ significance level respectively. These significances are post-trials, accounting for nine trials (one trial for each nightly pair tested). We note that this analysis does not search for evidence of variability at any timescales other than the nightly timescale. The flux differences present evidence for the TeV flux falling on a nightly timescale.  However, we did not observe an increase in TeV flux on this same short timescale. 

\begin{table}
 \centering
 \begin{tabular}{c|c|c|c}
\textbf{MJD}&  \textbf{Flux ($>$350 GeV)} & \textbf{p(F$_{1}$$>$F$_{2}$) ($\sigma$)} & \textbf{p(F$_{2}$$>$F$_{1}$) ($\sigma$)}   \\
& \textbf{$\times 10^{-12}$$\gamma$s cm$^{-2}$ s$^{-1}$} & & \\\hline
55911 & -1.5 $\pm$ 1.8 &  $<10^{-5}$ ($<0.1\sigma$)& \\
55912 & -1.02 $\pm$ 2.0 & &0.18 (0.23$\sigma$) \\\hline
\textbf{55918} & \textbf{13.5} $\pm$ \textbf{2.8} & \textbf{0.99} \textbf{(2.72$\sigma$)}& \\
55919 & 3.7 $\pm$ 1.3 & &$<10^{-5}$ ($<0.1\sigma$)\\\hline
55919 & 3.7 $\pm$ 1.3 & 0.76 (1.17$\sigma$) & \\
55920 & -0.77 $\pm$ 2.0 & &  $<10^{-5}$ $(<0.1\sigma$)\\\hline
55920 & -0.77 $\pm$ 2.0 & 3.9$\times$10$^{-3}$ ($<0.1\sigma$)\\
55921 & -1.02 $\pm$ 2.0 & & 7.6$\times$10$^{-4}$ ($<0.1\sigma$)\\\hline
55924 & 1.3$\pm$1.8 & $<10^{-5}$ ($<0.1\sigma$) & \\
55925 & 7.2$\pm$2.8 & &0.69 (1.01$\sigma$) \\\hline
55943 & 18.6$\pm$3.3 & 1.6$\times$10$^{-3}$ ($<0.1\sigma$) & \\ 
55944 & 18.6 $\pm$ 2.8 & &1.9$\times$10$^{-3}$ ($<0.1\sigma$) \\\hline
\textbf{55944} & \textbf{18.6} $\pm$ \textbf{2.8} &\textbf{ 0.99} \textbf{(3.57$\sigma$)} \\
55945 & 4.8 $\pm$ 2.1 & & $<10^{-5}$ ($<0.1\sigma$) \\\hline
55945 & 4.8 $\pm$ 2.1 & 7.4$\times$10$^{-3}$ ($<0.1\sigma$)\\
55946 & 4.1 $\pm$ 2.4 && 4.0$\times$10$^{-4}$ ($<0.1\sigma$)\\\hline
55946 & 4.1 $\pm$ 2.4 &  $<10^{-5}$ ($<0.1\sigma$)& \\
55947 & 13.7 $\pm$ 5.9 & &0.47 (0.63$\sigma$) \\\hline
\end{tabular}
\caption{The probabilities for both the flux increase and decrease per each pair of nightly separated fluxes. All probabilities shown are post-trials, accounting for nine trials (nine pairs of fluxes). All errors quoted are statistical only.}
\end{table}

\section{Multiwavelength Data}

\subsection{$\textit{Swift}$-XRT}
The \textit{Swift} X-ray Telescope (XRT) data \citep{Burrows} were reduced using the HEAsoft 6.12 package. Event files are calibrated and cleaned following the standard filtering criteria using the xrtpipeline task and applying the most recent Swift XRT calibration files. All data were taken in photon counting (PC) mode, with grades 0-12 selected over the energy range 0.3-10 keV. Since the count rate was below 0.5 counts s$^{-1}$ for all data, no evidence for photon pile-up in the core of the point-spread function (PSF) is evident. The source events are extracted from a circular region of radius of 30 pixels (47.2 arcsec). Background counts are extracted from a 40 pixel radius circle in a source-free region. Ancillary response files are generated using the xrtmkarf task, with corrections applied for the PSF losses and CCD defects. The latest response matrix from the XRT calibration files is applied. To ensure valid $\chi^{2}$ minimization statistics during spectral fitting, the extracted XRT energy spectra are rebinned to contain a minimum of 20 counts in each bin. Spectral analysis is performed with XSPEC 12.7. An absorbed power-law model, including the phabs model for the photoelectric absorption, is fit to each spectrum. A fixed column density is applied with an N$_{H}$ of 6.1 $\times 10^{21}$ cm$^{-2}$ \citep{Rea2010}. The spectral index of the source varied from -2.5 to -1.1 with reduced $\chi^{2}$ values ranging from 0.2 to 1.6. As observed in \citet{Smith2008}, the data show evidence for a correlation between the spectral index of the source and the 0.2-10 keV flux, with a Pearson correlation coefficient derived of 0.8$\pm$0.1.

The overall Swift-XRT light curve was extracted in the energy range of 2-10 keV and is shown in Figure 1. There were eight Swift-XRT observations that were taken simultaneously with VERITAS data (shown by grey bars in Figure 1).  In order to compare the VERITAS flux measurements with previous X-ray-TeV correlation studies, the $\>$350 GeV fluxes were interpolated to $\>$300 GeV fluxes using the fitted spectral index of -2.59 derived from the current observations. Both the VERITAS/Swift observations taken in 2011/2012 as well as archival VERITAS and MAGIC measurements \citep{Acciari2010} are shown in Figure 3. The correlation factor derived from the 2011/2012 observations was 0.36$\pm$0.32, consistent with two uncorrelated datasets. Including all simultaneous X-ray/TeV pointings from VERITAS and MAGIC results in a correlation coefficient of 0.33$\pm$0.14, which is consistent with no correlation.

\subsection{Fermi-LAT}

 $\textit{Fermi}$-LAT \citep{Atwood} analysis was performed on all available photons in the 0.3-300 GeV band obtained between December 1 2011 (MJD 55896)  and February 1 2012 (MJD 55958), in order to overlap as closely as possible with the VERITAS observations. The data were analyzed using Science Tools version v9r31p1, available from the Fermi Science Support Center (FSSC)\footnote[1]{http://fermi.gsfc.nasa.gov/ssc/}. Standard data quality cuts for Pass 7 event reconstruction were applied as recommended by the FSSC, with only ``source" (class 2) events being used for analysis.  Other standard cuts were also applied (e.g. zenith angle larger than 100$^{\circ}$ in order to reduce the contamination from atmospheric secondary gamma rays from near the Earth's limb \citep{Abdo2009b}). 

The LAT light curve was produced using the python likelihood tools and scripts available from the FSSC \footnote[2]{http://fermi.gsfc.nasa.gov/ssc/data/analysis/user/}. A region of interest (ROI) of 10$^{\circ}$ was chosen and a model file incorporating all 2FGL sources within a region of 15$^{\circ}$ was used for the initial fit. In this fit, all source showing a test statistic (TS) value of less than 1 for the data interval chosen were excluded. Additionally, all source more than 5$^{\circ}$ from the center of the ROI had fixed parameters in the model fitting. The resulting model was used to produce the daily binned lightcurve (shown in Figure 1), by fixing all 2FGL source model parameters (with the exception of LS I +61$^{\circ}$ 303 and the nearby pulsar 2FGL J0248.1+6021). To test for any correlation between the GeV and TeV flux, a correlation coefficient between the overlapping observations is calculated, with a coefficient of r=0.1$\pm$0.3, consistent with two uncorrelated datasets (see Figure 4). 

For spectral analysis, a binned maximum-likelihood method (\texttt{gtlike}) was used with an energy dependent ROI ranging from 2$^{\circ}$ to 10$^{\circ}$. In order to determine the background, the 2FGL catalog \citep{Fermi2cat} was used to account for the emission from all sources within a radius ranging between  3$^{\circ}$ to 15$^{\circ}$ (also a function of energy). The spectrum is satisfactorily fit (reduced $\chi^{2}$ value of 1.97 with 5 degrees of freedom) by a power law with exponential cutoff of the form A$\times$$\frac{E}{1 MeV}^{-\Gamma}$$\times$exp$^{-(E/E_{cutoff})}$, with A = (2.5 $\pm$ 0.9) $\times$10$^{-4}$ $\gamma$s MeV$^{-1}$ cm$^{-2}$ s$^{-1}$, $\Gamma$ = 2.13$\pm$0.06, and E$_{cutoff}$ = 3.98$\pm$0.42 GeV (see Figure 5). When comparing this spectrum to the one observed by VERITAS during contemporaneous observations, it is clear that the emission seen by Fermi-LAT experiences a dramatic fall off that is not observed in the TeV regime (see Figure 5). Since VERITAS observed the source at relatively large zenith angles (30$^{\circ}$-35$^{\circ}$), the energy threshold of the TeV observations do not allow for a detailed examination of the 100-200 GeV energy range.

\section{Summary and Discussion}

We have presented the results of a comprehensive multiwavelength campaign of the TeV binary LS I +61$^{\circ}$ 303 using VERITAS, Swift-XRT, and Fermi-LAT observations. The source was detected strongly in the TeV regime while not showing a significant correlation with the observed emission in the X-ray or  MeV-GeV regimes. The VERITAS differential energy spectrum obtained from these observations is well fit by a power law with spectral index consistent with previously published observations. The combination of the differential energy spectra obtained by both Fermi-LAT and VERITAS during the same time period reveals a puzzling lack of detected emission in the 1-200 GeV range. While the observation of this apparent discontinuity is not new (for example, \citet{Hadasch}) the previous GeV-TeV multi wavelength SEDs of LS I +61$^{\circ}$ 303 have, up until now, been constructed with data taken from various epochs. The observations detailed here represent the first time that a contemporaneous SED has been constructed with Fermi-LAT and IACTs since the launch of Fermi in 2008. The distinctive cutoff seen in the Fermi-LAT data, coupled with the significant detection of emission in the $>$200 GeV VERITAS energy range during the contemporaneous observations detailed in this work indicate that the observed emission in the Fermi-LAT/VERITAS energy ranges is produced by two separate populations of particles.  While we allow for the possibility that short term spectral variability in the Fermi-LAT energy regime could, in principle, produce a direct connection to the VERITAS TeV points, we consider such behavior unlikely given the spectral stability of the source in the Fermi-LAT regime \citep{Abdo2009}. 

 Given that the GeV spectral cutoff observed in LS I +61$^{\circ}$ 303 is strongly reminiscent of the typical cutoff shape seen in known Fermi-LAT pulsars, it is natural to suspect that emission in the system is indeed powered by an energetic pulsar. This model, as first proposed in \citet{MaraschiandTreves} and later developed and modeled in detail by \citep{Dubus},  explains the observed gamma-ray emission in LS I +61$^{\circ}$ 303 (as well as other known TeV binaries such as LS 5039 and PSR B1259-63) as being produced by the rotation power of a young pulsar. The inclusion of a pulsar in the system allows for much more flexibility in producing disparate populations of energetic particles (as appears to be observationally required in systems such as LS I +61$^{\circ}$ 303 and LS 5039) as there can be multiple acceleration regions for GeV/TeV energy particles: the inner pulsar magnetosphere, the shock interface between the pulsar and stellar winds (as well as multiple shocks separated by a contact discontinuity, as in \citet{Bednarek}),  acceleration within the pulsar wind zone (\citet{Sierpowski2008}), and potentially Coriolis effect generated shock fronts on scales much larger than the binary system (i.e. \citet{Zabalza}). 
 
 If we assume that the TeV emission is produced in the shock interaction between the two winds,  and that the GeV emission is produced in a second acceleration region or different seed particles, then it is possible that the GeV emission might be produced in the inner regions of the pulsar magnetosphere. The observed GeV variability could then be explained by absorption effects as the pulsar travels through the varying stellar wind density of the Be star. This would offer a natural explanation for the lack of GeV emission in the arguably similar TeV binary system HESS J0632+057; the pulsar beam in that system could be pointed away from our line of sight.  \citet{Bednarek} argues against the pulsar magnetosphere being the source of the GeV emission in the GeV/TeV binary systems, citing the lack of GeV emission from PSR B 1259-63 away from periastron where absorption effects should not play a strong role. This is indeed true and would necessitate a different mechanism for GeV emission in PSR B1259-63; however, given the relative uncertainty in the various physical parameters of the known TeV binaries, it is entirely possible that different mechanisms for emission could be at work in the different binary systems. 
 
The identification of LS I +61$^{\circ}$ 303 as a binary pulsar system is certainly not clear. For instance, despite many extensive searches \citet{radiosearch1,radiosearch2}, no pulsations have ever been detected, although it is possible that the dense stellar environment of LS I +61$^{\circ}$ 303 might preclude such a detection. 

The observations presented here also reveal the first strong evidence (99.97$\%$ confidence) for nightly variability in the source. If confirmed, this variability can provide crucial constraints on the size of the TeV emission region (i.e. the size of possible ``clumps" in the wind for pulsar binary models). Fast variability ($\sim$ second timescale) has already been associated with LS I +61$^{\circ}$ 303 in the X-ray regime \citep{Smith2008,TorresXray},  limiting the size of the X-ray emission region. Given the source strength of LS I +61$^{\circ}$ 303 and current sensitivity of IACT arrays, it is unlikely that such fast variability will be observed by the current generation of TeV instruments, even if occurring in the source. However, if the TeV and X-ray emission have a common mechanism, it could be possible to observe variability in the system on the order of tens of minutes during TeV flaring episodes.

These observations, taken in the context of past observations with VERITAS and MAGIC, also bring up the issue of the possible long-term variability seen in the system. Observations of this system with TeV instruments have only been taking place since 2006; while the observations have not been dense enough to make strong statements about the long term behavior of the source, it would appear that the source may go through a long-term modulation in the high energy regime.  The source was a strong TeV source in 2006/7 \citep{VERITASLSIDetection, Albert2006}, however, its TeV flux appears to have decreased over the succeeding years \citep{Acciari2010, Aleksic}. The ``normal" apastron TeV emission was markedly quiet, while the source was sporadically detected at near-periastron phases. The VERITAS observations taken in 2011/2012 indicate that the source may have returned to its ``normal" emission mode, with strong emission seen near apastron. Further long term observations of LS I +61$^{\circ}$ 303 with TeV instruments are key to understanding the possible multiyear modulation of the source and (given the lack of detected correlation between TeV emission and other bands) whether or not it is tied to similar emission modulation in radio \citep{Gregory2002}, X-ray \citep{LiXray, Chernya2012} and GeV gamma-rays \citep{Fermi2013}. 

\acknowledgments
This research is supported by grants from the U.S. Department of Energy Office of Science, the U.S. National Science Foundation and the Smithsonian Institution, by NSERC in Canada, by Science Foundation Ireland (SFI 10/RFP/AST2748) and by STFC in the U.K. We acknowledge the excellent work of the technical support staff at the Fred Lawrence Whipple Observatory and at the collaborating institutions in the construction and operation of the instrument.  We thank the Swift Team for scheduling contemporaneous observations and providing data and analysis tools. The authors would also like to thank Jeremy Perkins for his tireless assistance with Fermi-LAT data analysis.

\bibliographystyle{apj}
\bibliography{refs}

\pagebreak

\begin{figure}
\begin{center}
   \includegraphics[width=\textwidth,height=110mm]{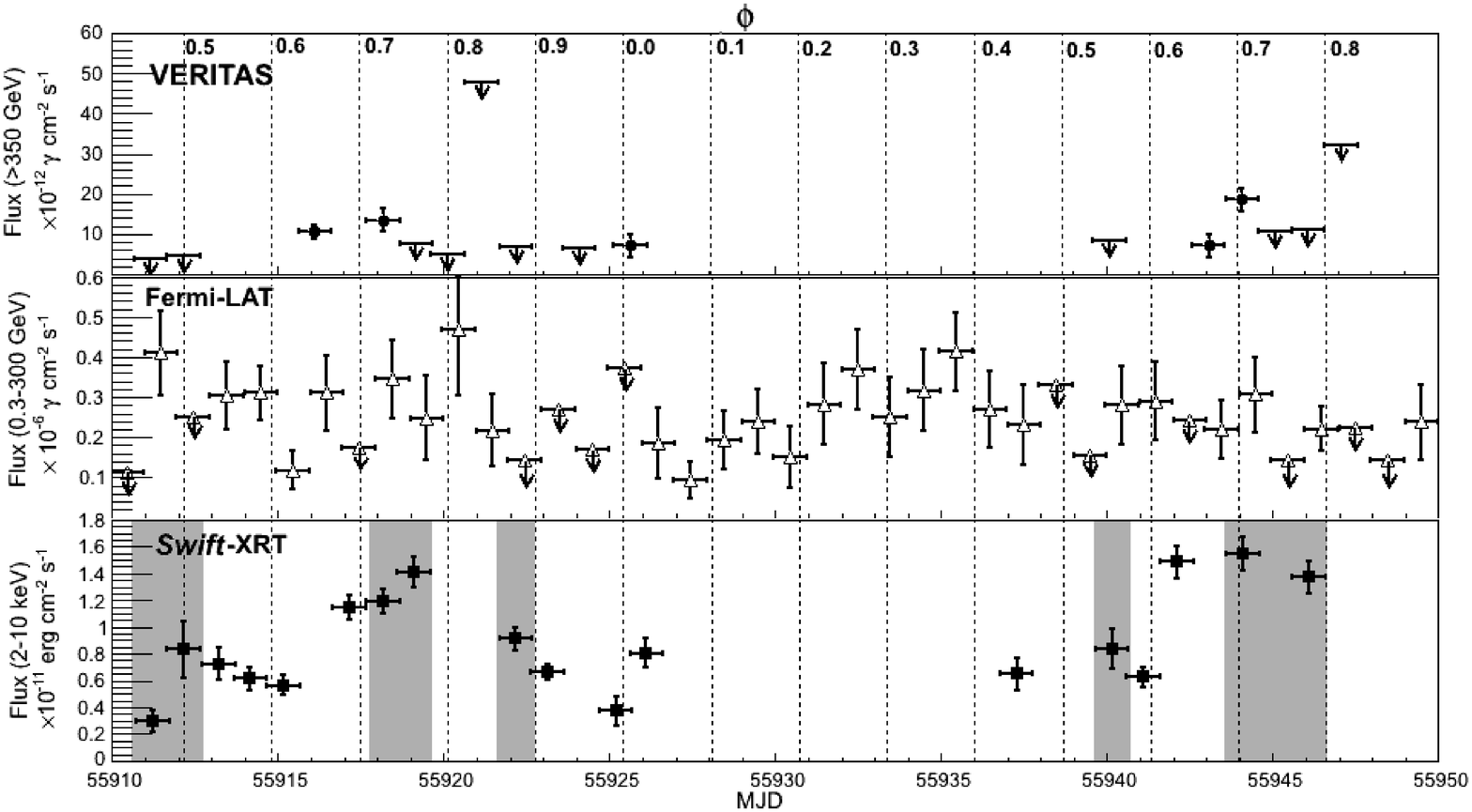}
\end{center}
\caption[]{The VERITAS ($>$350 GeV daily integrations, top), Fermi-LAT (0.3-300 GeV, middle), and Swift-$\textit{XRT}$ (0.3-10 keV, bottom) light curves for LS I +61$^{\circ}$ 303 during December 2011 - February 2012. The data is also shown as a function of orbital phase ($\phi$). VERITAS 99$\%$ flux upper limits are shown for points with $<3\sigma$ significance and are represented by arrows. Fermi-LAT upper limits (90$\%$ confidence level) are also shown by arrows. The grey shaded regions represent the observations obtained simultaneously which are used for the X-ray/TeV correlation studies in this work. }
\end{figure}

\begin{figure}
\begin{center}
   \includegraphics[width=\textwidth,height=110mm]{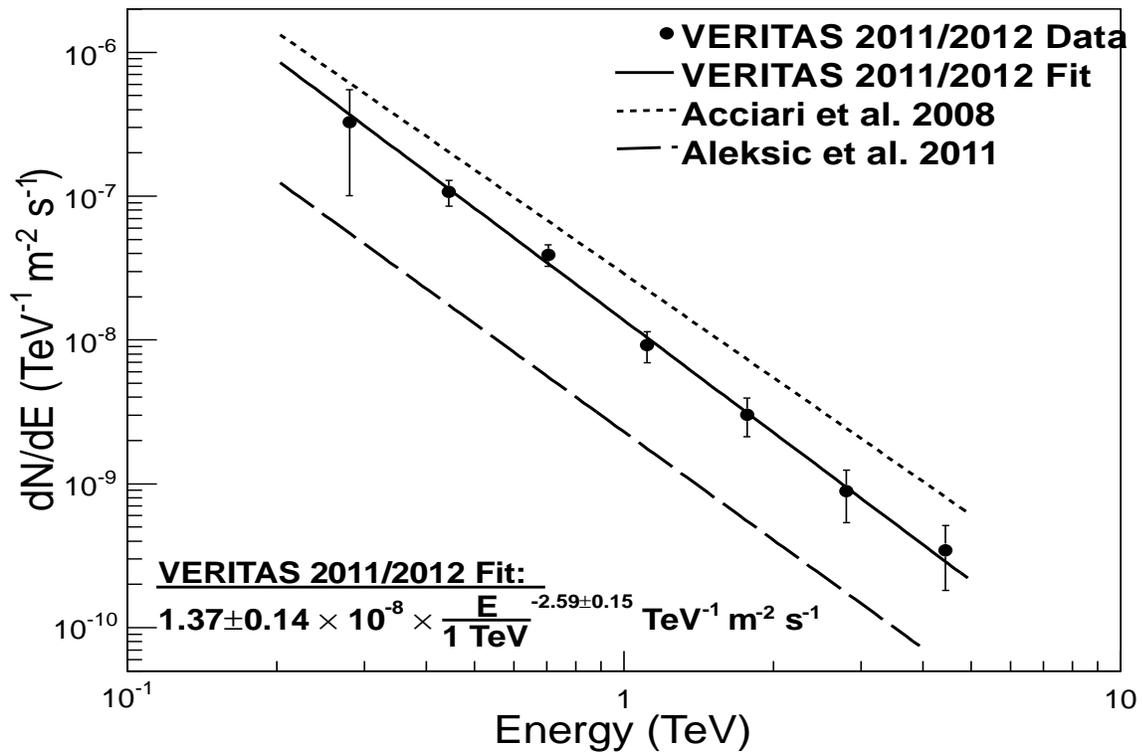}
\end{center}
\caption[]{ The VERITAS SED obtained from the 2011/2012 observations. We also show the SED power-law fit to both higher \citep{VERITASLSIDetection} and lower \citep{Aleksic} flux states of the source. }
\end{figure}

\begin{figure}
\begin{center}
   \includegraphics[width=\textwidth,height=110mm]{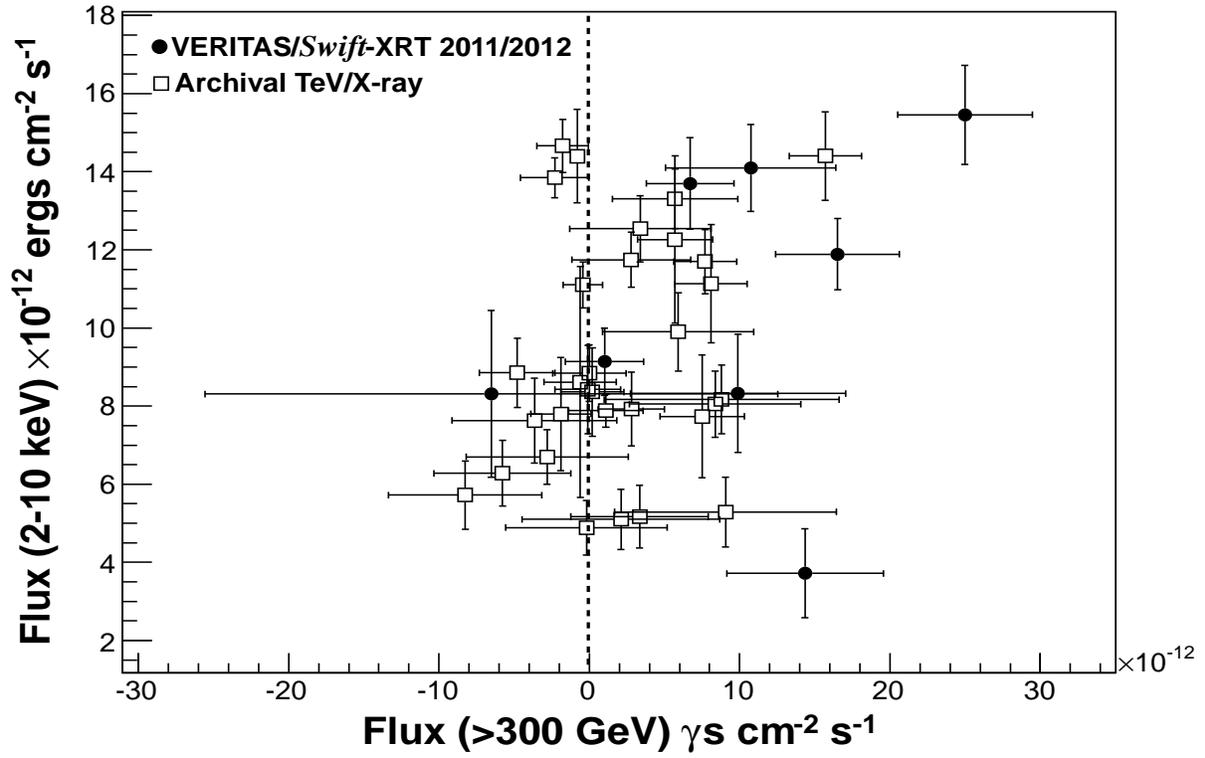}
\end{center}
\caption[]{ The comparison of the strictly simultaneous Swift-XRT and VERITAS data points. The data shows a correlation coefficient of 0.36$\pm$0.32, consistent with two uncorrelated data sets. }
\end{figure}

\begin{figure}
\begin{center}
   \includegraphics[width=\textwidth,height=110mm]{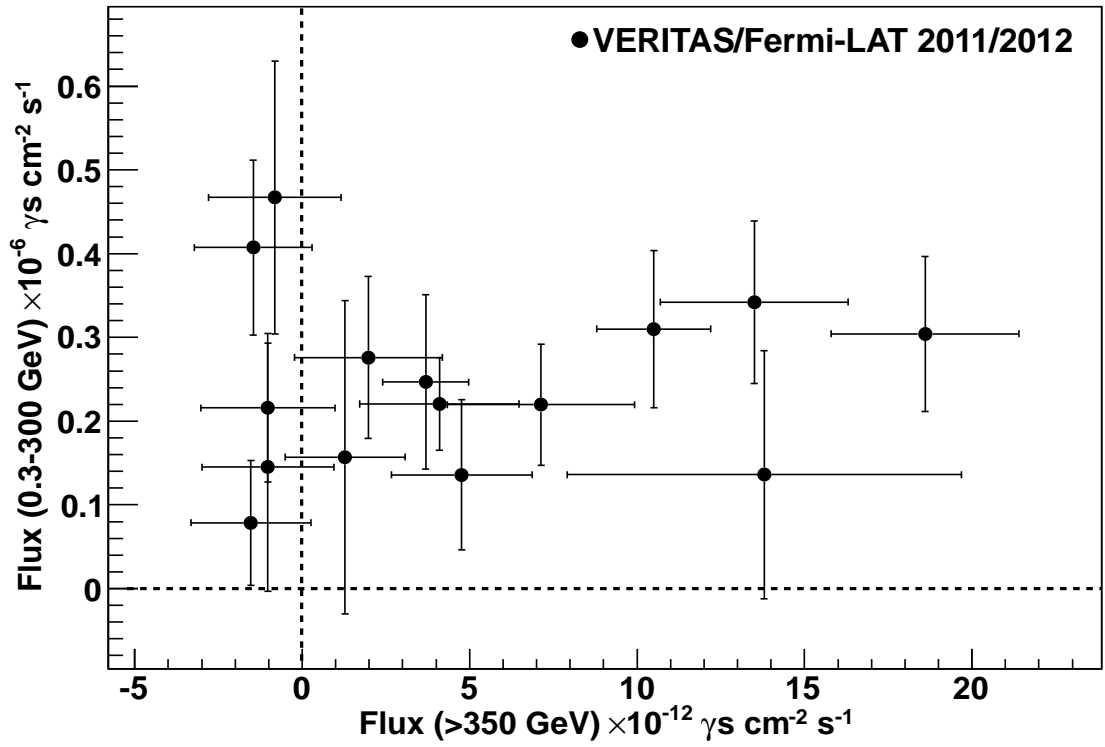}
\end{center}
\caption[]{ The comparison of nightly VERITAS and Fermi-LAT flux points. Analysis of the data results in a correlation coefficient of 0.1$\pm$0.3, consistent with two independent data sets. }
\end{figure}

\begin{figure}
\begin{center}
   \includegraphics[width=\textwidth,height=110mm]{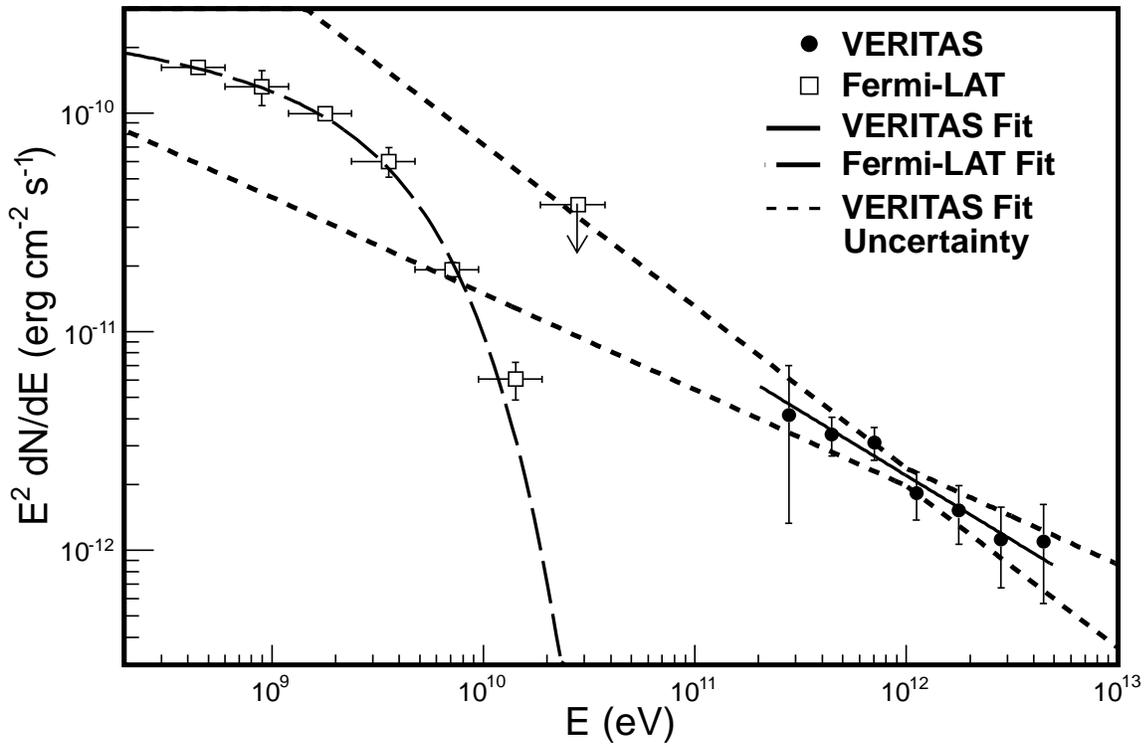}
\end{center}
\caption[]{ The VERITAS and Fermi-LAT spectral energy distribution. }
\end{figure}

\end{document}